\documentclass{ws-ijmpd}
\usepackage[super,compress]{cite}
\usepackage{graphicx,hyperref}

\begin{document}

\markboth{M. Giordano, A.~A. Nucita, F. De Paolis, and G. Ingrosso}
{Timing Analysis in Microlensing}

%
\catchline{}{}{}{}{}
%

\title{TIMING ANALYSIS IN MICROLENSING}

\author{Mos\`{e} Giordano\({}^{*}\), Achille A. Nucita\({}^{\dagger}\),
  Francesco De Paolis\({}^{\ddagger}\) and Gabriele Ingrosso\({}^{\S}\)}

\address{Dipartimento di Matematica e Fisica ``\emph{E. De Giorgi}'', \\
  Universit\`{a} del Salento, Via per Arnesano, CP 193, I-73100 Lecce, Italy \\
  INFN, Sezione di Lecce, Via per Arnesano, CP 193, I-73100 Lecce, Italy \\
  \({}^{*}\)mose.giordano@le.infn.it\\
  \({}^{\dagger}\)achille.nucita@le.infn.it\\
  \({}^{\ddagger}\)francesco.depaolis@le.infn.it\\
  \({}^{\S}\)gabriele.ingrosso@le.infn.it
}

\maketitle

\begin{history}
\received{Day Month Year}
\revised{Day Month Year}
\end{history}

\begin{abstract}
  Timing analysis is a powerful tool used to determine periodic features of
  physical phenomena.  Here we review two applications of timing analysis to
  gravitational microlensing events.  The first one, in particular cases, allows
  the estimation of the orbital period of binary lenses, which in turn enables
  the breaking of degeneracies.  The second one is a method to measure the
  rotation period of the lensed star by observing signatures due to stellar
  spots on its surface.
\end{abstract}


\ccode{PACS numbers: 95.75De, 97.10.Qh}

\section{Introduction}
\label{sec:introduction}

The timing analysis is the study of a repetitive feature of a phenomenon in
order to investigate its dynamic properties and infer its period.  Timing
analysis has always been a key tool for the search and the study of exoplanets.
For example, the radial velocity method is based on timing analysis of the
velocity profile along the line of sight of the parent star.  The transit method
relies on the periodic repetition of a same feature, the dimming of the light of
a star produced by the passing of a planet over the star's disk.  Furthermore,
cold and hotspots are generally present on the star surface, and if a starspot
is observed in different transits, it is possible to assess the rotation period
and velocity of the star, the sky-projected orbital obliquity of the planetary
system, and, in some cases, the true orbital
obliquity.\cite{2016arXiv160803095S}

Small variations from a perfectly periodic behavior give often hints about the
presence of perturbations that can reveal new phenomena.  For example, the first
exoplanets, around the pulsar PSR B1257+12, were discovered through the
detection of anomalies in the spin period of the neutron
star.\cite{1992Natur.355..145W} It has been
proposed\cite{2002ApJ...564.1019M,2005Sci...307.1288H,2005MNRAS.359..567A} to
perform timing analysis of transit times in order to find deviations from the
linear ephemeris (TTV, Timing Transit Variations), likely due to the presence of
an unknown body, as a moon or a planet.\cite{2009MNRAS.392..181K} This method
has been used, for example, to confirm and characterize the planet
Kepler-9\,d\cite{2010Sci...330...51H}.

Gravitational microlensing is an inefficient method to find new planets with
respect to transit and radial velocity methods, as a consequence of its low
optical depth, of the order of \(10^{-7}\) in the most favorable situations.
Yet, this technique, which is in some sense complementary to the other methods,
is the only able to detect very distant exoplanets in the Galactic Bulge or
outer galaxies\cite{2007A&A...469..115C,2009MNRAS.399..219I} and leads to a
complete characterization of the these systems.  Its use brought important
discoveries in the field of exoplanets, like the facts that cold
Neptunes\cite{2006ApJ...644L..37G}, free-floating
planets,\cite{2011Natur.473..349S} and terrestrial planets in low-mass
binaries\cite{2015ApJ...812...47U} are more common than previously expected.
Differently from most of the other methods to detect exoplanets, microlensing
offers generally a single non-repeatable event (except for self-lensing events,
like KOI-3278\cite{2014Sci...344..275K}).  Nonetheless, periodic deviations from
the standard microlensing amplification curves may give some information on the
lensing system or the source star.


We consider here two applications of timing analysis in the context of
gravitational microlensing.  The first one, discussed in
Section~\ref{sec:rotating-binaries}, allows us to estimate, with some caveats,
the orbital period of the binary lens.\cite{2014MNRAS.438.2466N} The second
application, outlined in Section~\ref{sec:starsp-rotat-source}, is about the
determination of rotating period of the source star thanks to the presence of
stellar spots on its surface.\cite{2015MNRAS.453.2017G}

\section{Rotating Binaries}
\label{sec:rotating-binaries}

The discovery of exoplanets by microlensing is made possible by the fact that
the presence of a companion of the main lensing object induces deviations from
the standard Paczy\'{n}ski curve which characterizes a single lens
event.\cite{2016Univ....2....6D}

In order to fit the amplification curve due to a binary lens system a number of
parameters has to be taken into account.  Four parameters are those used to
model a single lens event: \(u_{0}\), the distance of closest approach between
the lens and the source projected in the sky; the time \(t_{0}\) of closest
approach; the Einstein time \(t_{\textup{E}}\); the angle \(\theta\) of the
source's trajectory in the plane of the lens.  Further parameters are the
projected separation \(s\) between the lenses, and their mass ratio
\(q = m_{2}/m_{1}\), being \(m_{1}\) and \(m_{2}\) the masses of the two lenses.
If the background source is not point-like, finite size effects should be taken
into account by considering the source radius \(\rho_{*}\) as well.
Furthermore, if the orbital motion of the lenses is not negligible, six
additional parameters are required to model the event: the semi-major axis \(a\)
(that is related to the lenses sepration \(s\)), the orbital eccentricity \(e\),
the time of passage at periapsis \(t_{\textup{p}}\), the inclination angle \(i\)
between the normal to the lens orbital plane and the line of sight, the
orientation \(\varphi\) of the orbital plane in the sky, and the orbiting versus
(either clockwise or counterclockwise).  This is only one of the possible sets
of parameters that can be used to describe such events.  Furthermore, there are
some degeneracies that may add complexity to this scheme.  For example, under
certain circumstances the caustics resulting from two lenses separated by a
projected distance \(s\) or \(1/s\) have the same structure, implying that the
observed light curves will appear the same.  This is the so called
\emph{close-wide degeneracy} that may occur in lensing systems with \(q \ll 1\),
like in planetary systems.\cite{1998ApJ...500...37G}

The modeling of a microlensing event by binary lenses can be a daunting task,
due to the large parameter space to be explored, and could take several CPU
hours.  One of the most delicate aspects of a fitting procedure, that can speed
up the convergence to the best solution, is the choice of good starting values,
as close as possible to the true ones.  Most of the parameters of a
Paczy\'{n}ski curve are often easy to estimate by eye looking to the
amplification curve: \(t_{\textup{E}}\) is the timescale of the event, the
half-width of the curve is roughly \(t_{\textup{E}}\); \(t_{0}\) is the position
of the main peak of the curve; the asymptotic behavior of the amplification
\(A\) for small values of the distance of closest approach
is\cite{2016ApJ...823..120N} \(A \sim 1/u_{0}\), so that \(u_{0}\) can be
readily guessed by measuring the amplification of the main peak.  It has been
argued\cite{1992ApJ...396..104G} that also other parameters of a binary lens
event can be estimated by eye in some cases, but not all.

We also mention that also the source may be in a binary system, even if binary
sources have been seldom recognized in photometric microlensing observations.
In this respect, a more efficient method for the detection of their orbital
motion is given by astrometric microlensing.\cite{2016ApJ...823..120N}

\subsection{Estimating the orbital period}
\label{sec:estim-orbit-peri}

The orbital period \(P\) of a binary lensing system is related to its semi-major
axis \(a\) by the Kepler's third law:
\begin{equation}
  P = 2\pi\sqrt{\frac{a^{3}}{Gm_{1}(1 + q)}}
\end{equation}
Thus, measuring the orbital period \(P\) of the lensing system by means of
timing analysis, independently from a fit to the microlensing amplification
curve, could give a quantity that is important by itself and help in better
constraining some of the free parameters of the full fit of the amplification
curve.

The method we first proposed in Ref.~\refcite{2014MNRAS.438.2466N} consists in
searching for the amplification curve by a single lens that fits best the
amplification curve by the binary lens under examination, which shows
non-negligible effects of the lens orbital motion.  Of course, this method can
be applied only when the shape of this curve closely resembles that of a
Paczy\'{n}ski curve.  The residuals between the two curves present periodic
features that allow the estimation of the orbital period of the binary lensing
system by using timing analysis.  This procedure removes the underlying trend
and lets the periodic features of the curve emerge.  This is a simple task to
perform because the parameters of a Paczy\'{n}ski curve can be easily guessed,
as discussed above.  Another limitation is that, in order to be able to apply
the timing analysis, the orbital period of the lenses should be shorter than the
Einstein time of the event, or we must have a long observational window so that
the lenses complete at least two orbits during the observations.  In addition,
care must be used when analyzing the results, because this procedure may give
half of the true period, even though the order of magnitude is correctly
retrieved.

\subsection{Simulations of binary lens events}
\label{sec:simul-binary-lens}

We tested this method by performing simulations of microlensing events involving
binary lenses with relevant orbital motion during the event.  There are
different techniques to calculate the amplification by binary lenses, each one
with its advantages and drawbacks.  Some of these techniques are:
\begin{itemize}
\item Witt \& Mao method:\cite{1995ApJ...447L.105W} these authors showed that,
  if the source is far enough from the caustic, the amplification of a
  point-like source by binary lenses can be computed by solving a fifth-order
  polynomial with complex coefficients.  This method is fairly fast, the major
  bottleneck being the search of the roots of the
  polynomial,\cite{2010MNRAS.408.2188B} but the restriction on the position of
  the source (far enough from the caustic) limits its application.
\item Hexadecapole method:\cite{2008ApJ...681.1593G} this method enables the
  calculation of the amplification of an extended source and stems from the
  fourth-order Taylor expansion of this quantity.  It consists in measuring the
  amplification of a point-like source in thirteen positions.  For the
  amplification by binary lenses the Witt \& Mao method can be employed to
  compute these amplifications of a point-like source.  This method is fast, is
  designed for extended sources, works with any lens configuration and takes
  into account the limb-darkening effect as well, but can be applied far enough
  from the caustic (but generally closer than the Witt \& Mao method).
\item Inverse ray shooting:\cite{1986A&A...164..237S,1986A&A...166...36K} it
  consists in ``shooting'' a large number of ``photons'' from the observer
  towards the source star and apply the lens equation\cite{2016Univ....2....6D}
  to compute the deflection by the lenses.  This method is accurate and works
  with extended source and any lens configuration, even when the source crosses
  the caustic, the locus of the points in the lens plane where the amplification
  diverges to infinity.  On the other hand, this technique is also
  computationally costly, especially in the case of lens orbital motion.
\end{itemize}
We adopted a hybrid approach, using all techniques but each one only where
necessary, in order to take the best of all methods and speed up calculations.
In particular, we decided to use the inverse ray shooting method when the source
is close to the caustics, and the hexadecapole method combined with the Witt \&
Mao amplification\footnote{In order to expedite the search for the roots of the
  polynomial we used the fast General Complex Polynomial Root
  Solver\cite{2012arXiv1203.1034S} as implemented in the Julia package
  \texttt{PolynomialRoots.jl}, which is free and open source.  More information
  can be found at \url{https://github.com/giordano/PolynomialRoots.jl}.} when
the source is far enough from the caustic.  We address the reader to
Ref.~\refcite{2014MNRAS.438.2466N} for details.

\begin{figure}
  \centering
  \includegraphics[width=0.85\textwidth]{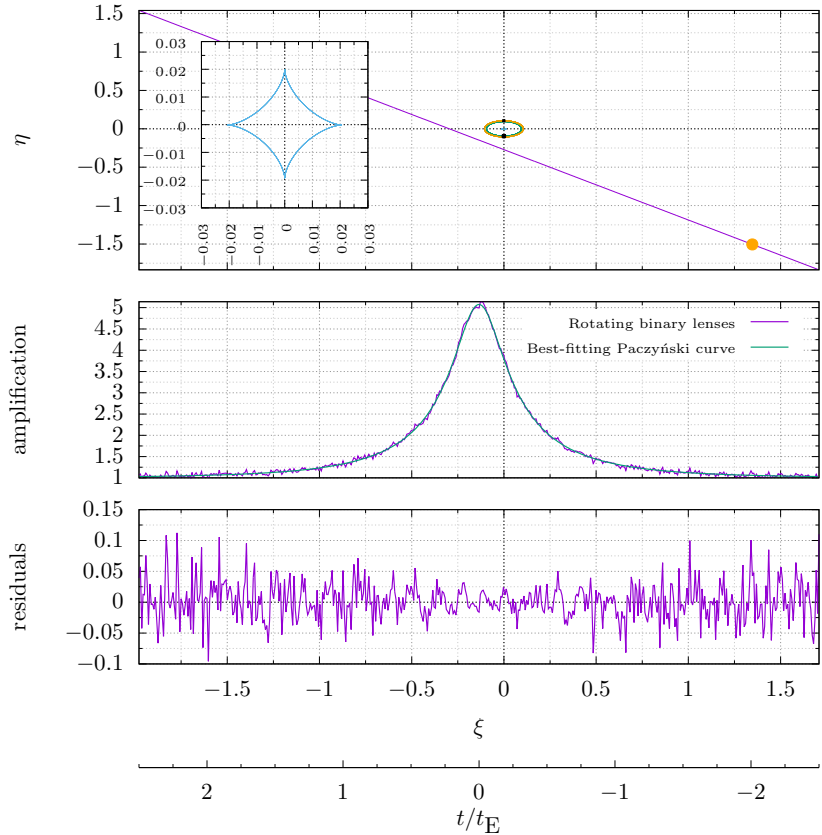}
  \caption{\emph{Upper panel}: Geometry of the microlensing event.  Purple
    straight line represents the trajectory of the source (the yellow dot on
    bottom right) in the lens plane \(\xi-\eta\), from bottom right to top left.
    The small green and yellow elliptical curves around the origin are the
    orbits of the lenses.  The inset on the left shows a zoom in of the central
    caustic of the lenses, at the end of the simulation.  \emph{Middle panel}:
    Amplification of the source by the rotating binary lenses
    (\(A_{\textup{BL}}\)) and its best-fitting Paczy\'{n}ski model
    (\(A_{\textup{SL}}\)).  Gaussian noise has been added to the simulated
    amplification \(A_{\textup{BL}}\).  \emph{Lower panel}: residuals
    \((A_{\textup{BL}} - A_{\textup{SL}})/A_{\textup{SL}}\) of the amplification
    curve by rotating binary lenses relative to the Paczy\'{n}ski model.  Note
    these are not the absolute residuals, that would be
    \(A_{\textup{BL}} - A_{\textup{SL}}\).}
  \label{fig:rotating-binary}
\end{figure}
Here we present a new simulation of a microlensing with rotating binary lenses.
We adopted dimensionless quantities, with lengths expressed in units of the
Einstein radius \(R_{\textup{E}}\) and times in units of the Einstein time
\(t_{\textup{E}}\).\cite{2014MNRAS.438.2466N} The source star has radius
\(\rho_{*} = 0.03\), limb-darkening coefficient \(\Gamma = 0.5\), and its
projected distance of closest approach to the center of mass of the lenses is
\(u_{0} = 0.2\).  The lenses have mass ratio \(q = 0.9\), their orbit has
semi-major axis \(a = 0.2\), eccentricity \(e = 0\), and period \(P = 0.4\).
Inclination angle \(i\) and rotation \(\phi\) are both set to \(0\).  The
geometry of the system and the results of the simulations are shown in
Fig.~\ref{fig:rotating-binary}.

\begin{figure}
  \centering
  \includegraphics[width=0.8\textwidth]{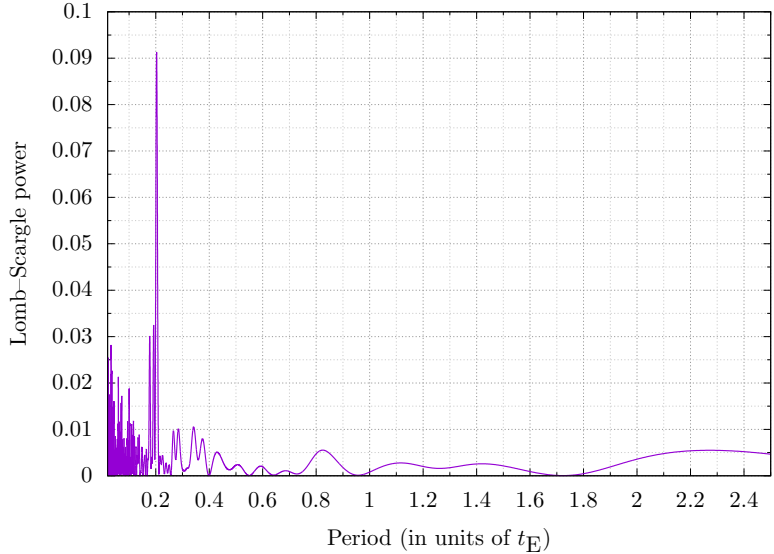}
  \caption{Lomb--Scargle periodogram of the residuals in
    Fig.~\ref{fig:rotating-binary}.  We adopted the standard normalization,
    suggested by
    Refs.~\protect\refcite{1976Ap&SS..39..447L,2009A&A...496..577Z}, for which
    the Lomb--Scargle power \(p(P)\) is in the range \([0, 1]\), standing
    \(p(P) = 0\) for no periodicity in the data with period \(P\), and
    \(p(P) = 1\) for a ``perfect'' periodicity with period \(P\).}
  \label{fig:rotating-binary-periodogram}
\end{figure}
We then applied the generalized Lomb--Scargle
periodogram\cite{2009A&A...496..577Z} to the relative residuals shown in
Fig.~\ref{fig:rotating-binary}, using the \texttt{LombScargle.jl}
package\footnote{The \texttt{LombScargle.jl} package is free and open source,
  more information about its installation and use can be found at
  \url{https://github.com/giordano/LombScargle.jl}.}  written in Julia language.
We found that spurious peaks in the residuals appears within the central part of
the event,\cite{2014MNRAS.438.2466N} so we had to remove a small region around
\(t = 0\) before actually performing the periodogram.  The result of this timing
analysis is shown in Fig.~\ref{fig:rotating-binary-periodogram}.  The period
with the highest peak is \(P = 0.203\), which is half of the simulated orbital
period \(P = 0.4\) of the lenses, because the caustic is
symmetric\cite{2014MNRAS.438.2466N} (see inset of
Fig.~\ref{fig:rotating-binary}). In Ref.~\refcite{2014MNRAS.438.2466N} are
presented cases where the caustic is not symmetric and the correct period is
found.

\section{Starspots on Rotating Source Star}
\label{sec:starsp-rotat-source}

Many authors suggested to exploit microlensing events in order to study
irregularities on the surface of the source star, like cold- and hotspots,
either with
photometric\cite{2000ApJ...529...69H,2000MNRAS.316..665H,2002MNRAS.335..195C,2002MNRAS.335..539H}
or polarimetric microlensing observations.\cite{2015MNRAS.452.2587S} This effect
is particularly important since the presence of stellar spots on the source may
fake planetary features in events that are actually due to a single lens, as
occurred in the case of MOA-2010-BLG-523.\cite{2013ApJ...763..141G} However,
none of the above mentioned works took into account the possibility for the
source star and/or the binary lens system to rotate.  We filled this gap in
Ref.~\refcite{2015MNRAS.453.2017G} and showed that also in this case timing
analysis could be a powerful tool to get more information on the considered
system.

\subsection{Application of timing analysis}
\label{sec:appl-timing-analys}

In a manner similar to the one described in Section~\ref{sec:estim-orbit-peri},
we will show that by applying timing analysis to the residuals of a fit of a
``static'' curve to an intrinsically periodic curve we can retrieve the rotating
period of the lensed star.

In Ref.~\refcite{2015MNRAS.453.2017G} we performed simulations of microlensing
events with single and binary lenses, involving a rotating source and/or
rotating lenses.  The amplification \(A_{\textup{s}}\) of the spotted star has
been calculated as the weighted average of the amplification \(A(\vec{r})\) over
the star disc \(\mathcal{S}\), using as weight the surface brightness
\(f(\vec{r})\)
\begin{equation}
  A_{\textup{s}} = \frac{\int_{\mathcal{S}}A(\vec{r})
    f(\vec{r})\mathrm{d}^{2}\vec{r}}{\int_{\mathcal{S}}f(\vec{r})\mathrm{d}^{2}\vec{r}}.
\end{equation}
The amplification \(A(\vec{r})\) is the amplification of a point-like source by
either a single or a binary lens, depending on the event.  Also in this work we
employed a hybrid approach for calculating the amplification by a binary lens,
like the one discussed in Section~\ref{sec:simul-binary-lens}, using inverse ray
shooting when the source is close to the caustic, and Witt \& Mao method far
enough from the caustic.  The surface brightness profile \(f(\vec{r})\) is
defined as
\begin{equation}
  f(\vec{r}) =
  \begin{cases}
    l(\vec{r})   & \text{outside the spot} \\
    c~l(\vec{r}) & \text{inside the spot},
  \end{cases}
\end{equation}
where \(l(\vec{r})\) is the brightness of the unspotted star, and \(c\) is the
\emph{contrast parameter}, that is the ratio between the brightness of the spot
and the unspotted surface.  The case \(c>1\) corresponds to hotspots, \(c<1\) is
for coldspots.

\begin{figure}
  \centering
  \includegraphics[width=0.85\textwidth]{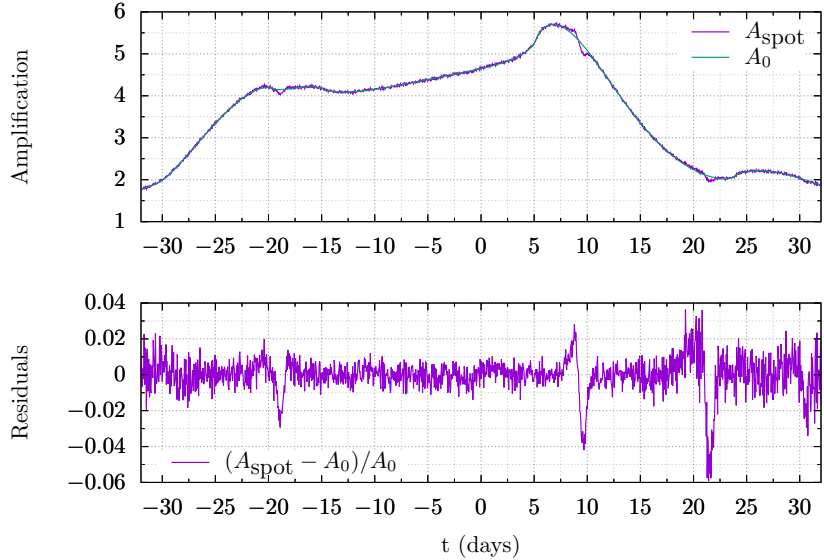}
  \caption{Results of the simulation described in the text.  \emph{Upper panel}:
    amplification curve of the unspotted star (\(A_{0}\)) and the spotted one
    (\(A_{\textup{spot}}\)).  Gaussian noise has been added to the amplification
    \(A_{\textup{spot}}\) of the spotted star.  \emph{Lower panel}: the
    residuals \((A_{\textup{spot}} - A_0)/A_0\) of the spotted star model
    relative to the unspotted one are represented.  Note these are not the
    absolute residuals, that would be \(A_{\textup{spot}} - A_0\).}
  \label{fig:rotating-spot}
\end{figure}
Here we present a new simulation of a microlensing event of a rotating star by a
static binary lens.  The lens system is constituted by two equal mass objects,
so their mass ratio is \(q = 1\), and they are at a projected separation of
\(b = 1\).  The ratio \(D_{\textup{l}}/D_{\textup{s}}\) between the
observer-lens distance \(D_{\textup{l}}\) and observer-source distance
\(D_{\textup{s}}\) is equal to \(0.5\).  The source star has rotation period
\(P = 10~\mathrm{d}\), typical of main-sequence G2 stars observed by
\emph{Kepler},\cite{2013A&A...557L..10N} and its projected radius is
\(\rho_{*} = 0.1\).  The limb-darkening coefficient adopted in the simulation is
\(\Gamma = 0.45\).  We considered both a star with a spot on its surface and a
star without any such feature.  Starspots can have a variety of sizes, covering
up to \(22\%\) of the stellar hemisphere.\cite{2009A&ARv..17..251S} We simulated
a coldspot with center on the equator (corresponding to colatitude
\(\theta = \pi/2\)), contrast \(c = 0.1\), and radius equal to \(0.2~\rho_{*}\).
In Fig.~\ref{fig:rotating-spot} the results of the simulations are shown.
\begin{figure}
  \centering
  \includegraphics[width=0.8\textwidth]{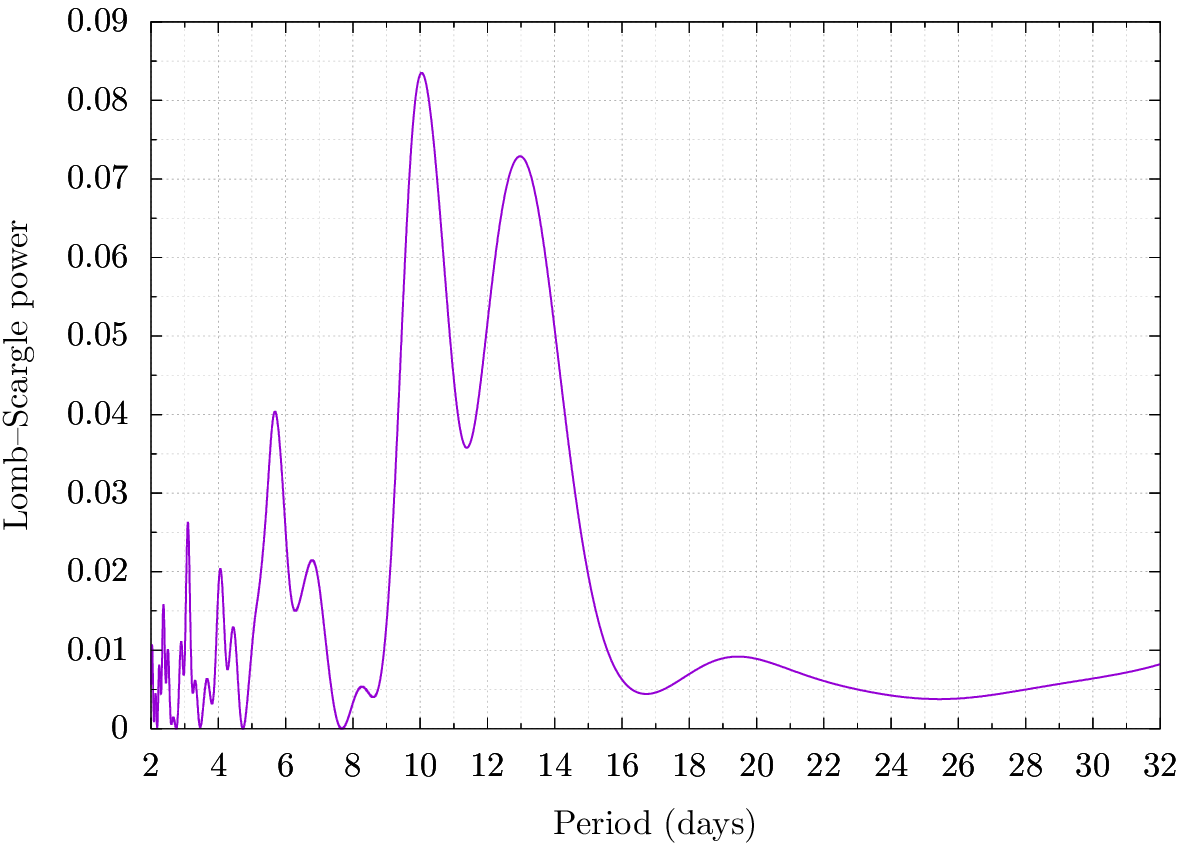}
  \caption{Lomb--Scargle periodogram of the residuals in
    Fig.~\ref{fig:rotating-spot}.  It has the same standard normalization used
    in for the periodogram in Fig.~\ref{fig:rotating-binary-periodogram}.  The
    simulation spans 64 days, therefore the largest period we consider in the
    periodogram is 32 days, the half width of the time series.}
  \label{fig:lomb-scargle-spot}
\end{figure}
We applied the generalized Lomb--Scargle periodogram to the absolute value of
the relative residuals shown in Fig.~\ref{fig:rotating-spot}, using the fast
algorithm\cite{1989ApJ...338..277P} provided by the \texttt{LombScargle.jl}
package.  In Fig.~\ref{fig:lomb-scargle-spot} the periodogram is shown.  The
period with the largest power is \(P_{1} = 10.05~\mathrm{d}\).  Another peak is
present at \(P_{2} = 12.97~\mathrm{d}\), but while the highest peak at \(P_{1}\)
is replicated by a broad peak at the higher harmonic \(P = 19.44~\mathrm{d}\),
there is no higher harmonic peak for the secondary peak at \(P_{2}\), suggesting
that the period at \(P_{1}\) is the real one, which is indeed very close to the
rotation period \(P = 10~\mathrm{d}\) of the source star we used in the
simulation.  Thus, in this case we showed that the application of the timing
analysis to the residuals allows us to measure the rotation period of the source
star.

\section{Comments}
\label{sec:comments}

In this paper we considered two possible applications of timing analysis to
gravitational microlensing events.  In both cases, the use of this tool enabled
us to recover important information about the simulated system.

In Section~\ref{sec:rotating-binaries} we showed that even applying an
inappropriate fitting function to the amplification curve can be useful.  We
fitted a microlensing event due to a rotating binary with the Paczy\'{n}ski
model, with the mere purpose of enhancing the periodic features through a sort
of detrending procedure.  Using the Lomb--Scargle periodogram with the
residuals, we were able to find half of the true orbital period of the binary
lensing system with good accuracy.  Nevertheless, the knowledge of at least the
order of magnitude of the orbital period makes it possible to better constrain
some of the free parameters of the real fit to the observed curve.

The presence of spots on the surface of the star induces small deviations from
the standard amplification curve.  In Section~\ref{sec:starsp-rotat-source} we
have been able to get the rotation period of the star by measuring the residuals
of the amplification of the spotted star relative to the amplification of the
unspotted star, and then applying the Lomb--Scargle periodogram on them.  The
analysis of spot features requires high-precision and high-cadence photometry,
because these signatures can last a few hours.  The Korean Microlensing
Telescope Network (KMTNet) can collect up to 6 data points per
hour\cite{2014ApJ...794...52H} and could be able to observe starspots features.
By using observational parameters of OGLE-III and OGLE-IV campaigns, we
estimated in Ref.~\refcite{2015MNRAS.453.2017G} the number of events with these
features to be about \(4.7\) per year towards the Galactic Bulge.

\section*{Acknowledgments}

We acknowledge the support by the INFN project TAsP.

\appendix

\bibliographystyle{ws-ijmpd}
\bibliography{bibliography}

\end{document}